# Volumetric Helical Additive Manufacturing

Antoine Boniface [1], Florian Maître [1], Jorge Madrid-Wolff [1], Christophe Moser [1, *]

[1] Laboratory of Applied Photonics Devices, School of Engineering, Ecole Polytechnique Fédérale de Lausanne, CH-1015 Lausanne, Switzerland

Corresponding author: christophe.moser@epfl.ch

**Abstract**

**3D printing has revolutionized the manufacturing of volumetric components and structures for various fields. Thanks to the advent of photocurable resins, several fully volumetric light-based techniques have been recently developed to overcome the current limitations of 3D printing. Although fast, this new generation of printers cannot fabricate objects whose size typically exceeds the centimeter without severely affecting the final resolution. Based on tomographic volumetric additive manufacturing, we propose a method for volumetric helical additive manufacturing (VHAM) multi-cm scale structures without magnifying the projected patterns. It consists of illuminating the photoresist while the latter follows a helical motion. This movement allows for increasing the height of the printable object. Additionally, we off-center the modulator used for projecting the light patterns to double the object's lateral size. We demonstrate experimentally the interest in using these two tricks for printing larger objects (up to 3 cm × 3 cm × 5 cm) while maintaining high resolution (< 200 µm) and short print time (< 10 min).**

## 1. Introduction

Over the last decade, 3D printing technologies have experienced unprecedented developments and changes. They now enable fabricating complex volumetric objects rapidly and inexpensively. This makes 3D printers especially attractive and pertinent for various fields including the aerospace industry or medical applications [1,2]. Until recently, the paradigm in light-based 3D printing or additive manufacturing (AM) mainly relied on using a vat of liquid photopolymer resin, out of which the object is constructed sequentially, layer by layer or voxel by voxel[3]. An ultraviolet (UV) light beam cures the resin one layer at a time whilst a platform moves the object being made downwards after each new layer is hardened. The UV light is either raster scanned, which solidifies the resin point by point as in stereolithography (SLA)[4], or flashed onto the resin curing the whole layer at once as in digital light processing (DLP) technologies[5,6]. Due to the layer-by-layer nature of the printing process, these light-based additive manufacturing techniques are subject to major geometric constraints and throughput limitations.

The past few years have seen the emergence of several fully volumetric additive manufacturing (VAM) technologies that move away from the layer-by-layer approach. Two-photon photopolymerization represents the state-of-the-art of volumetric printing with light[7]. It enables the fabrication of microscale objects with a lateral resolution of 100 nm and axial resolution of 300 nm but is slow, with a printing speed of just 1–20 mm$^3$/h, and requires expensive femtosecond laser sources. Recently, teams have demonstrated that two-photon lithography can be sped up through spatiotemporally focusing of ultrafast lasers[8,9]; while two-step absorption, instead of two-photon absorption, has enabled microfabrication without the need for pulsed lasers[10,11]. For high-speed solidification of centimeter-scale objects, two volumetric approaches have been recently developed. In one, called xolography, polymerization is locally induced at the intersection of two light beams of different wavelengths using photoswitchable photoinitiators[12]. It is a fast technique that can print building blocks of only tens of microns but which requires special photoinitiators and is strictly limited to optically clear and low-absorptive resins. In the second approach, coined computed axial lithography (CAL) or tomographic volumetric additive manufacturing, an entire three-dimensional object is simultaneously solidified by irradiating a liquid photo-sensitive resin volume from multiple angles with dynamic 2D light patterns[13–15]. These 2D light patterns are calculated from the target 3D object using 3D-to-2D transforms, like the Radon transform, similarly to X-ray Computer Tomography Scans[16]. In a tomographic printer, a Digital Micromirror Device (DMD), which offers millions of degrees of freedom to spatially modulate the light intensity of the input beam, is used to produce the 2D patterns. By projecting patterned light into the liquid resin from multiple angles, a 3D energy dose is accumulated. In the regions where such energy dose exceeds a polymerization threshold, solidification occurs. The object is finally printed when all target voxels inside the liquid precursor receive an irradiation dose above this said threshold[17]. Thanks to high-power laser diodes, the tomographic photopolymerization of cm-scale objects can be performed as fast as within 30 to 120 seconds with resolutions of around 100 μm[14]. After printing, the surrounding liquid unpolymerized resin can be washed away to reveal the desired solid printed part.

Using this process, it was first possible to obtain cm-scale objects with a resolution of 300 μm (ref. [13]). When the polymerization starts, the refractive index of the resin changes, locally perturbing the propagation of light. This intricate phenomenon was not taken into account initially for computing the 2D light patterns, limiting print fidelity. It has been shown that re-calculation of the patterns based on measured feedback from a first sacrificial print combined with an optimized low-étendue illumination can bring the final print resolution to 80 μm (for positive features) and 500 μm (for negative features)[15]. However, print resolution and fidelity are still limited, mainly by three primary sources of error: *(1)* the projected light patterns; *(2)* the chemical and optical properties of the resin (*e.g.*, polymerization diffusion, absorbance or scattering, and change in refractive index upon polymerization), and *(3)* the optical projection setup. The goodness of the algorithm is essential to produce patterns whose back projections form a volumetric dose of light that best represents the target object inside the resin. Several recent works have been proposed to optimize the set of patterns with respect to *(i)* the target dose by including physics priors[18,19] or *(ii)* more appropriate loss functions that enhance contrast[20]. Regarding the resin's opacity, it is clear that optically transparent materials allow for propagating sharp patterns of high fidelity, but the same patterns would be inevitably blurred and thus lose their finest features in scattering media. The detrimental effect of scattering can be reduced through refinements to the calculated tomographic patterns[18] or through refractive-index matching of the resin[21]. The chemical diffusion of free radicals may also cause unwanted polymerization that enlarges the print resolution by a few microns[22,23], but this effect can be mitigated by doping the resin with free-radical quenchers[24,25]. Ultimately, the optical resolution of the printer dictates the achievable printed voxel size. In DLP and tomographic VAM, optical resolution is determined at best by the features of the modulator used to pattern light, namely the DMD. Here we use a DLP7000 chip from Texas Instrument that has on its surface $N_x \times N_y = 768 \times 1024$ micro-mirrors (pitch = 13.6 μm) arranged in a rectangular array capable of displaying 8-bit images.

In our optical system, the DMD image is magnified by a factor of 1.66; the resulting pattern onto the vial is 1.74 cm × 2.33 cm with a resolution of 23 μm (see a detailed sketch of the experimental setup in Supplementary S.1). The only way to increase the size of the printed objects without compromising the resolution is to move the DMD with respect to the vial or vice versa. Inspired by spiral CT (ref. [26]), we propose to move the sample around the light beam with a helical trajectory. Additionally, we show that lateral printable size can be doubled without compromising resolution by off-centering the optical axis with respect to the rotation axis of the photoresist vat (Fig. 1.a). Together, these two tricks increase the number of building blocks inside the vial by a factor of up to 12. Here these available printed voxels are used to print larger objects up to 3 cm × 3 cm × 5 cm in a few minutes without compromising the printing resolution (< 200 μm).

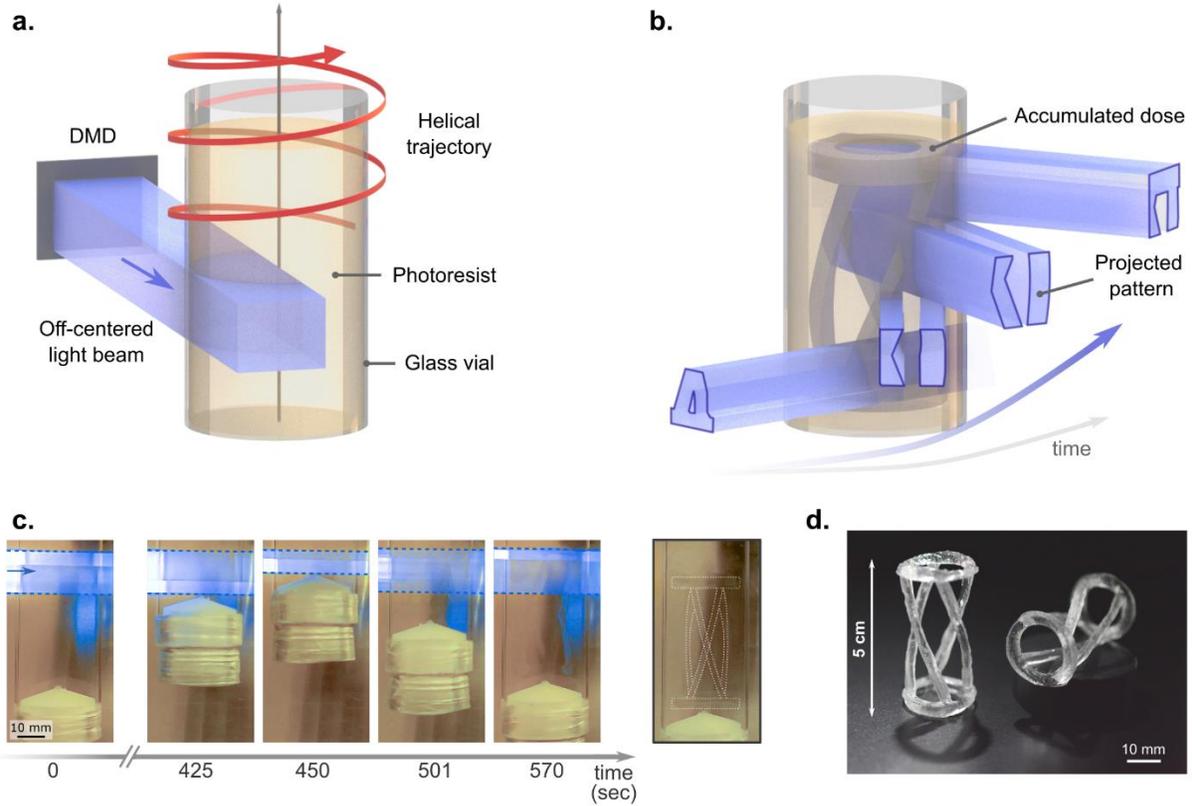

*Figure 1: Principle of tomographic volumetric helical additive manufacturing (VHAM). **a.** Simplified schematic of the helical printer. A laser beam (λ = 405 nm) is modulated in intensity with a DMD before propagating through the photoresist. The vial containing the latter is off-centered, such that one lateral edge of the rectangular beam intersects the center of the cylindrical container. The vial is placed in rotation and move continuously up and down defining a helical trajectory. **b.** Schematic representation of the helical printing procedure. A series of patterns of light projected over multiple angles trigger at different time the polymerization. At a given time only a subpart of the resin is exposed to light. In fact, it is the helical movement that ensures the solidification of the whole object. **c.** Time lapse. Two cycles up/down and 12 turns were necessary to solidify the resin in the desired geometry (see 3D model in b.). This is achieved in less than 10 min. The rotation stage holds the vial by the bottom. A white cap prevents resin's leakage to the outside. **d.** Final part obtained after washing out the unpolymerized resin. Scalebar: 10 mm.*

The principle of tomographic volumetric helical additive manufacturing (VHAM) is given in figure 1.a. We combine a rotation and a translation stage to set the glass vial (diameter = 32 mm) containing the photoresist in a helical motion. We must emphasize here that all the resin is not illuminated at once as in conventional tomographic VAM. Here, in VHAM, the whole resin is entirely excited only after one complete cycle comprising a bottom-up and a top-down pass. Half a cycle (only up or down) includes α rotations of the vial. As the vial follows a helical trajectory, α also represents the number of patterns stacked along the vertical axis. There are some overlapping regions between the patterns so that after a turn its lower and upper parts coincide. The size of the overlap is fine-tuned by adjusting the vial's rotation speed to the vertical movement of the translation stage; which is essential to ensure continuity of the printed objects. A detailed workflow regarding the computation of the VHAM patterns is provided in Methods and fig. S.2. In this work, the rotation speed is between 8 and 10 $°.s^{-1}$ which respectively gives a vertical linear speed of 458 $\mu m.s^{-1}$ and 366 $\mu m.s^{-1}$. After a few up and down cycles, the light dose accumulated inside the resin at different heights and over multiple angles is sufficient to solidify it as shown on the schematic figure 1.b. This usually happens after 2 or 3 vertical cycles and is in general completed in less than ten minutes (fig. 1.c). Note that because of light absorption, patterns projected at θ and θ+180° do not irradiate the volume of resin in the same way. We take this into account by performing a blank half turn (simply put, no projection and no vertical translation over 180°) between two vertical cycles. It results that for 3D structures with no central symmetry, the number of projected patterns doubles. To give an idea, one may need to send around 10,000 patterns for printing with α = 3 and an angular resolution of 0.18°.

The final 3D printed structure is obtained after some post-processing including a washing and post-curing steps as described in Methods. For this particular helical tower structure α = 3, which means 4α = 12 times more printable voxels inside the resin compared to conventional tomographic VAM.

## 2. Results

We report on the capabilities of helical tomographic VAM through a series of different printed structures in transparent acrylics. The photo-curable resin used in this work is prepared by combining a commercial polyacrylate (PRO21905 from Sartomer), with 0.6 mM phenylbis (2,4,6-trimethylbenzoyl) phosphine oxide as a photoinitiator (TPO from Sigma Aldrich) in a planetary mixer. The optical transparency together with its stiffness and ease of use makes it a candidate of choice to present our printer's competences in terms of object size and resolution. In figure 3, we show the prints of five different 3D models, relatively large (2 cm wide at least), with different heights. For all of them, the DMD is off-centered with respect to the vial's rotation axis, but the parameter α is adjusted to fit the height of each object. As in conventional tomographic VAM, these complex and hollow geometries are printed in a short time (3-10 min) without the need for support structures.

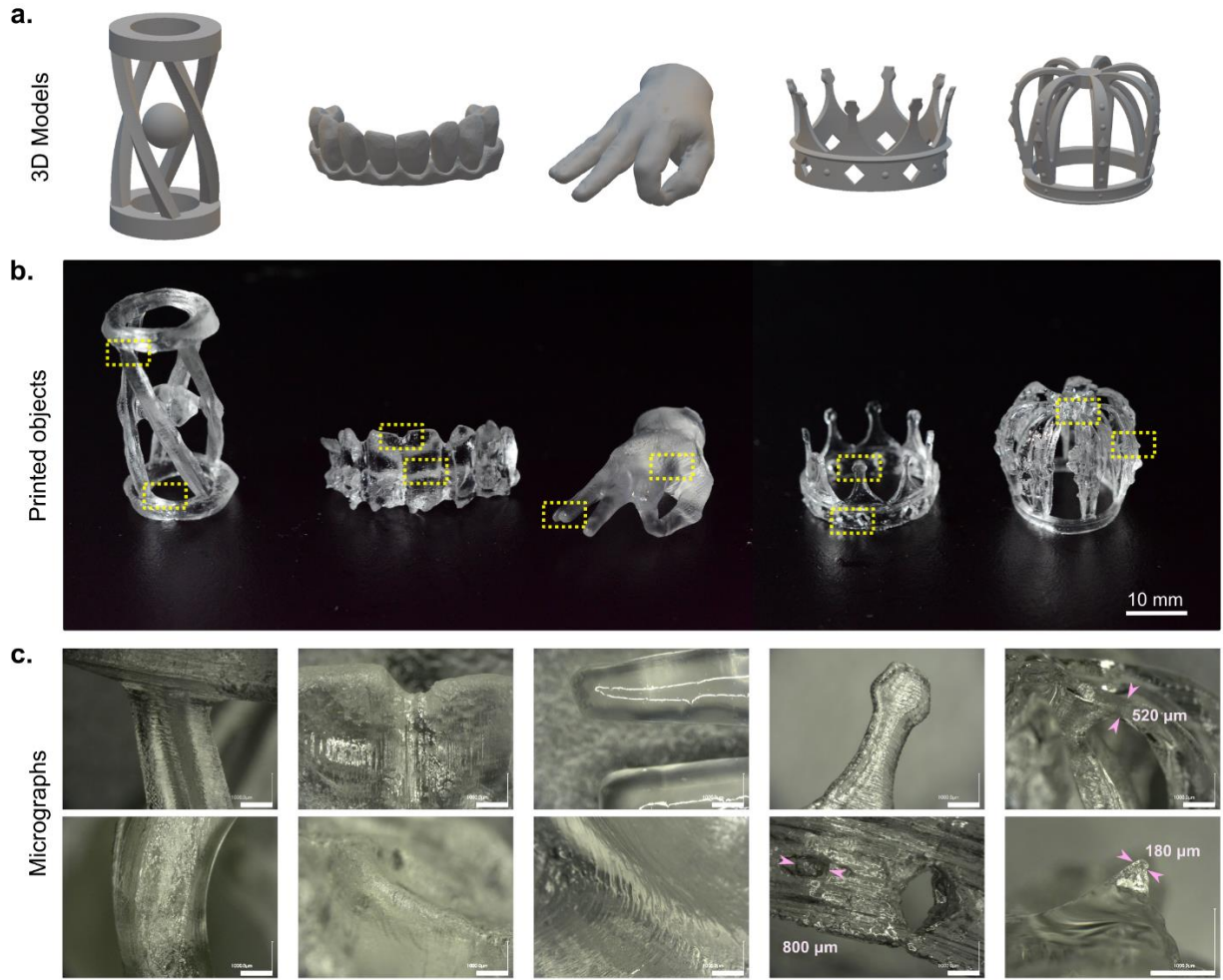

*Figure 2: Examples of 3D printed objects using tomographic VHAM. **a.** 3D model. **b.** Photographs of the obtained prints. **c.** Microscopic images to better appreciate the printed fine details. Scalebars: **b.** = 10 mm, **c.** = 1 mm.*

The absence of layering offers excellent surface quality as one can see in the micrographs in Figure 2.c, especially for the prints of the hand and the teeth. Striations, similar in appearance to a few tens of microns thick layers, can be observed. They are caused by a self-induced waveguide effect, driven by the gelation material nonlinearity and can potentially be removed[27]. Suspended complex structures as thin as the 520 μm double arches of the crown and its 180 μm spikes in relief can be printed with high fidelity. The printer is not only capable to print sharp edges like the square pillars of the helical tower but also round and curved surfaces like the circular base of the structure. We also report on the possibility of printing in a few minutes optically clear models of teeth with high resolution, as shown in Figure 3. The speed of the process and the achieved level of details might make helical tomographic VAM particularly interesting for dental industry[28]. Other works have demonstrated that tomographic VAM can be used to fabricate polymer-derived SiOC ceramics[29] or silica glass from nanoparticle-loaded acrylic resins[30].

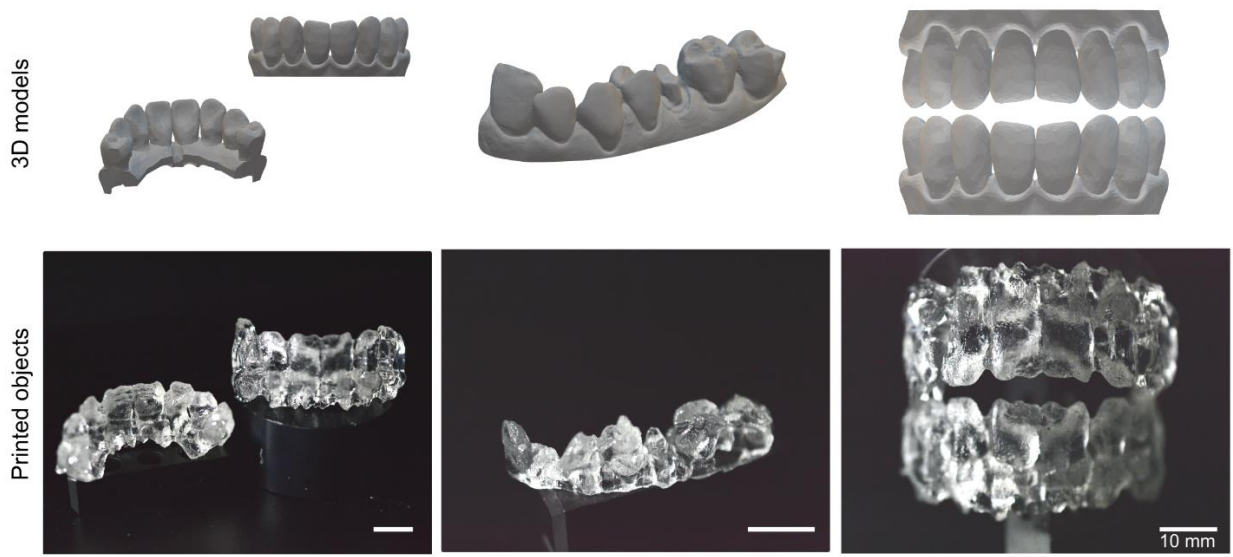

*Figure 3: High-resolution printing of adult-size teeth. Scalebars: 10 mm.*

### 3. Discussion and Conclusion

Having a look at conventional tomographic VAM, one rapidly foresees the advantages related to the helical motion. Usually in tomographic VAM, the available $N_x \times N_y$ pixels of the DMD illuminate the whole rotating volume of resin at once, as represented in Figure 4.a. In theory this configuration provides $\pi/4 \; N_x^2 \times N_y$ independent printable voxels. Off-centering the DMD doubles the lateral resolution or increases the number of printed voxels by a factor 4. Additionally, the continuous translation of the vial gives the possibility of printing taller objects (see Figure 2.b). The increase in the vertical direction is described by the parameter α. Further comparisons against other light-based volumetric additive manufacturing methods are presented in table 1 of the supplementary material.

In this work, we printed objects with α up to 3, meaning objects 3 times taller. Overall, this new optical configuration increases the number of printed voxels inside the vial by a factor up to 4α = 12 compared to conventional tomographic VAM, see Figure 4.b. This size increase is at the cost of the printing speed since the resin must be exposed to light for a longer time to reach the threshold dose. It is possible to lower the threshold dose by increasing the concentration of photoinitiator, but one must be careful not to make the resin too absorptive, or by increasing the laser power.

As in any light-based VAM methods, printing larger structures comes at least with two challenges. On the one hand, light has to penetrate deeper inside the resin to cure the whole volume; light absorption is therefore more critical. On the other hand, light has to propagate straight over longer distances. Whatever the printing technology, light is never perfectly collimated and divergence can be at the origin of strong deviation between the model used for computing the patterns and the real experiment[15,19]. In our case, we deliberately chose to reduce the numerical aperture of our optical projection system to reduce the beam divergence in order to preserve the printing fidelity although it is at the cost of the resolution and printing speed since we cut part (the high spatial frequencies) of the incoming light. We provide in Supplementary characterization of the beam divergence in our VHAM printer and the effect of the numerical aperture (Supplementary S.3), a numerical model to simulate the beam divergence (Supplementary S.4), and two studies based on this model to see the effect of both divergence (Supplementary S.5) and defocusing (Supplementary S.6) on the resulting light dose. Although divergence inevitably affects print resolution, we successfully printed features size of 200 μm for objects as big as 2.5 cm × 2.5 cm × 3 cm (see in Figure 2.c).

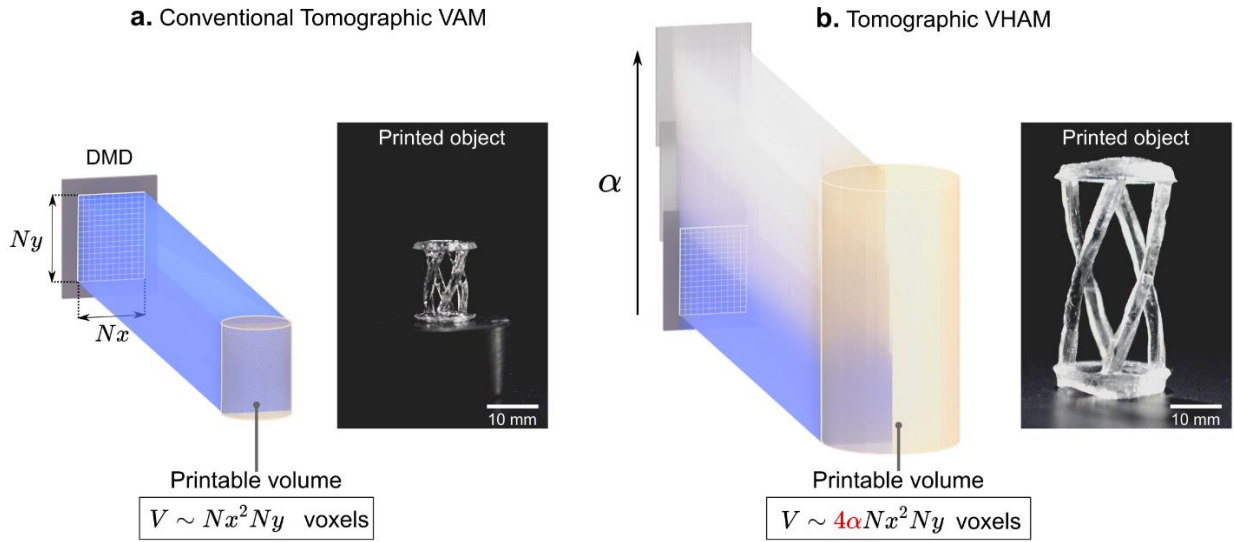

*Figure 4: Increasing the number of printed voxel generated from the same DMD. **a.** Conventional tomographic volumetric additive manufacturing. **b.** Tomographic volumetric helical additive manufacturing. Scalebars: 10 mm.*

We expect similar resolution for objects taller than 3 cm (for a greater α) since the vertical motion should not affect resolution. For comparison, we calculate that a modified version of CAL -one using an LED instead of a laser diode- with the corresponding magnification would probably provide a resolution around 700 μm for objects of the same size. To further improve the printing resolution it could be interesting to use a better model for computing the light patterns (for instance by including the divergence of the beam) or getting feedback from the printing process[15,31]. The chemical system also plays an important, or rather essential role concerning the obtained resolution. In particular, the spatial resolution can be improved from the previously reported value, by intentionally introducing radical quenchers in the resin besides oxygen that naturally acts as inhibitor of free-radical polymerization (*e.g.* TEMPO). In terms of printing time, helical VAM is comparable with xolography for similar object sizes. Both of them seem to offer very competitive capabilities but their working principles remain different, and depending on the target structure or material used one might offer better results than the other.

In summary, we have presented a new light-based technique for volumetric printing of multi-centimeter scale objects. It builds up on tomographic VAM to significantly increase (up to a factor 12) the number of printable voxels while keeping the same modulator device for projection and without compromising too much the printing resolution. This was achieved by off-centering the modulator and translating continuously the resin around the patterned light beam. These simple modifications can be easily made on existing tomographic printers and opens up new possibilities for high-resolution and high-speed fabrication of objects whose size up to 3 cm × 3 cm × 6 cm. Helical tomographic VAM might be therefore appealing for applications in fields where cm-scale objects must be manufactured individually, such as in the dental industry.

## 4. Materials and Methods

### 4.1 Optical Setup

The optical setup for tomographic VHAM is presented in figure S.1. Two 405 nm laser diodes, with a combined nominal power of 1.8 W, are collimated and combined into a single beam with a D-shaped mirror. The combined beam is then coupled into a square-core optical fiber (CeramOptec WF 70x70/115/200/400N, core size 70 μm by 70 μm, numerical aperture 0.22), in order to spatially homogenize the beam from the two laser diodes. The outgoing square beam is then magnified to match the rectangular aperture of the DMD via an aspheric lens $L_3$ and a set of two orthogonal cylindrical lenses $L_4$ and $L_5$ for maximizing the light efficiency. Note that the cylindrical lenses have different focal lengths ($f_4$ = 250 mm and $f_5$ = 300 mm), in order to adjust the square beam from the fiber square output facet to the rectangular area of the DMD. The DMD suffers from diffractive effects due to the blazed grating formed by the micromirrors (pitch = 13.6 μm). This effect can cause a large fraction of the reflected light to be lost in diffracted orders depending on the incidence angle of the illumination beam. The surface of the DMD is imaged via a 4*f*-system into a cylindrical glass vial containing the photopolymerizable resin. In the Fourier plane (between $L_6$ and $L_7$), an iris blocks the unwanted diffraction orders from the DMD. This iris also effectively controls the numerical aperture of the beam, and its aperture can be adjusted to balance beam width and Rayleigh length. A refractive-index matching bath of vegetable oil is used to remove the lensing distortion caused by the cylindrical interface of the vial[32]. Compared to conventional tomographic VAM, the DMD is off-centered with respect to the vial's rotation axis and the resin can be moved vertically thanks to a linear stage (travel range of 10 cm). A side view camera placed perpendicular to the optical axis monitors the printing process. A red LED that does not influence the photopolymerization is used for this purpose.

### 4.2 Computation of the light patterns

The computation of the light patterns from the 3D model target dose relies essentially on the Radon transform as developed for tomographic imaging. However in the case of printing, the problem to solve is reversed: one has to compute the 2D patterns from the 3D dose whereas in imaging the algorithm aims at reconstructing the 3D object from a set of 2D measurements. As in tomographic VAM[13,33] the starting point is the 3D model, i.e. the object one intends to print. The latter is first voxelized into a three-dimensional binary matrix, where the entries "1" indicate the presence of matter and "0" its absence at each particular location in space. The voxel size depends on the optical setup and is in our case around 23 μm. The dimension of the matrix is therefore given by the target object size divided by the voxel size. This matrix also represents the normalized target dose that one would need to deposit in a transparent resin to polymerize it in the desired geometry. A series of dose projections over multiple angles are calculated from the Radon transform. More precisely, the patterns are obtained using a filtered back-projection algorithm followed by an optimization subject to positivity constraint. Please note that this forward model assumes the use of optically-clear materials, in which light propagates straight and without attenuation. The obtained patterns are too large to be entirely projected with the DMD. For this reasons they are cropped twice. First, along the horizontal axis, because the DMD is off-centered and second vertically to account for the up and down moving of the vial. These two crops allows for reducing the image size to a pattern that can be projected onto the DMD. To avoid any printing discontinuity along the vertical direction the patterns are softened on the corresponding edges with a smoothing mask that contains an overlapping region of adjustable height. The last step consists of padding with zeros the patterns to fit the DMD size.

**4.3 3D models**

We used 3D models that were freely available online: "Okay" hand by "cyclone" (https://www.thingiverse.com/cyclone), from (https://www.thingiverse.com/thing:10017/files, CC BY-SA 3.0 license).

The *.stl* files of the models were voxelized into three-dimensional matrices using the *stl*-to-voxel code of Christian Pederkoff (https://github.com/cpederkoff/stl-to-voxel). The voxel size was initially set be equal to one DMD mirror before calculating back-projection patterns. Note that the rotated 3D object constitutes a very large 4-dimensional matrix ($x$, $y$, $z$, $\theta$). When GPU memory (11 Gb, NVIDIA GeForce RTX 2080 Ti OC) was insufficient to accommodate it, we down-sized it spatially, calculated the back-projection patterns, and up-sized it to a scale of 1 voxel = 1 DMD mirror by interpolation.

**4.4. Hardware control**

In the optical setup, we used the python library from Jeffrey Mohan to display images on the DMD (https://gitlab.phys.ethz.ch/mohanj/holography/-/tree/master/dmd) and the Motion library from Zaber to control the rotational and translational stages (https://www.zaber.com/software/docs/motion-library/ascii/references/python/). We used a National Instruments DAQ card (PCIe-6321, X Series) as a reference clock to trigger the image display on the DMD.

**4.5 Photoresist & volumetric printing**

A liquid pentaacrylate commercial resin (PRO21905, Sartomer) was mixed with 0.6 mM phenylbis (2,4,6-trimethylbenzoyl) phosphine oxide as a photoinitiator (TPO, Sigma Aldrich) in a planetary mixer (Mazerustar Kurabo KK-250). The mix was poured into 32 mm cylindrical glass vials, and bubbles were removed by intermittent sonication for about an hour. Printing was performed at room temperature.

In order to rinse the printed objects after printing, they were gently taken from the vials with a spatula, immersed in 50mL falcon tubes containing isopropyl alcohol, and sonicated for several minutes at room temperature. Objects were then post-cured immersed in glycerol in a curing chamber (Form Cure, Formlabs) under UV light to remove the stickiness of their surface.

**4.6 Imaging**

Fabricated objects were were imaged with a DSLR camera (D3100, Nikon, Japan) with a f = 2.8 macro lens (AF-S Micro Nikkor 40 mm, Nikon), and a digital microscope (VHX-5000, Keyence, USA) with magnifications between 20 and 100×.

**Acknowledgements**

This project has received funding from the Swiss National Science Foundation under project number 196971 - "Light based Volumetric printing in scattering resins". The authors thank the free or open-source tools (and their contributors) which were used in this work, including Tinkercad.com, FreeCADweb.org, Inkscape.org, Python.org, PyTorch.org, and Thingiverse.com.

**Conflicts of interest**

C. Moser is a shareholder of Readily3D SA. A. Boniface, J. Madrid-Wolff, and C. Moser have filed a patent application for helical volumetric additive manufacturing.

# Volumetric Helical Additive Manufacturing: Supplementary Material

Antoine Boniface [1], Florian Maître [1], Jorge Madrid-Wolff [1], Christophe Moser [1]

[1] Laboratory of Applied Photonics Devices, School of Engineering, Ecole Polytechnique Fédérale de Lausanne, CH-1015 Lausanne, Switzerland

## S1. Experimental setup

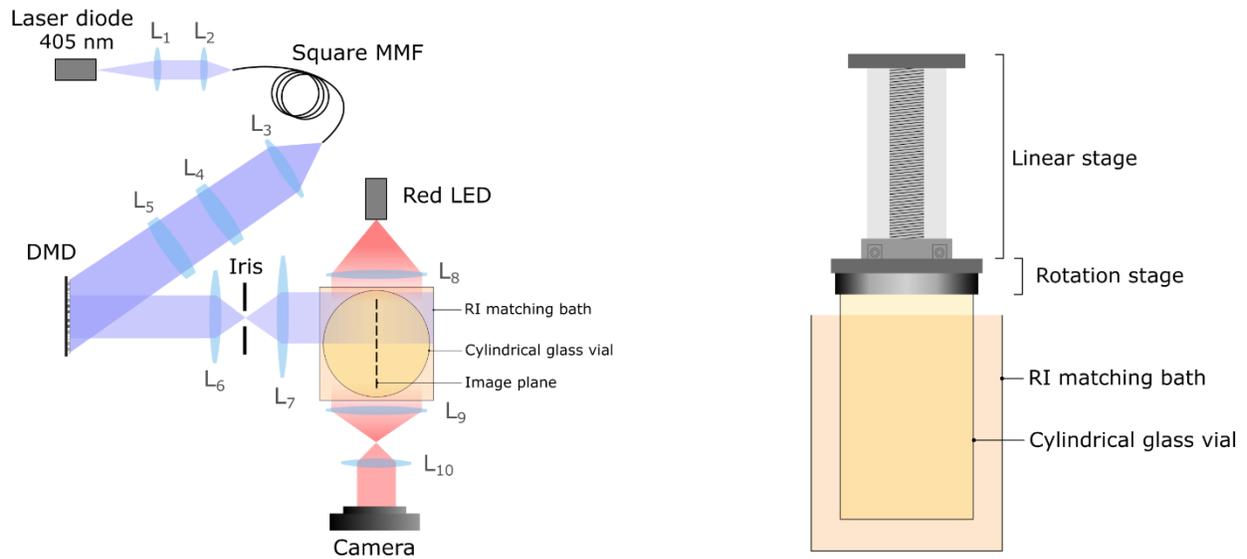

***Figure S.1: Experimental setup.*** *L1: f = 4.02 mm aspheric lens ; L2: f = 11 mm; aspheric lens; Square MMF: 70μm by 70μm square-core optical fiber; L3: f = 2 mm aspheric lens; L4: f = 250 mm cylindrical lens; L5 : f = 300 mm cylindrical lens; DMD: digital micromirror device (Vialux V-7000 VIS); L6: f = 150 mm lens; A: variable aperture; L7: f = 250 mm lens; cylindrical glass vial: outer diameter 32 mm, wall thickness 1 mm; L8: f = 40 mm lens; L9: f = 250 mm lens; L10: f = 35 mm lens; camera: DCC3260C USB 3.0 CMOS Camera, Thorlabs; Linear stage: LSQ075A, Zaber ; Rotation stage: X-RSW60C, Zaber.*

## S2. Computation of the light patterns

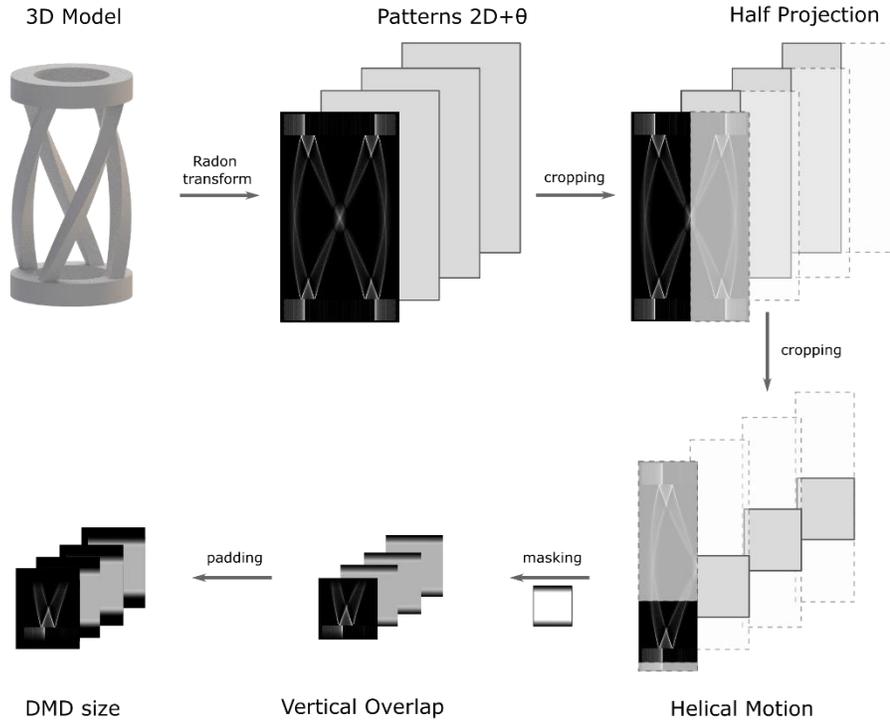

*Figure S.2: Workflow for computing the light patterns*

## S3. Characterization of the Beam Divergence

The beam divergence of a laser beam is a measure of how fast the beam expands apart from the imaging/focal plane. Note that it is not a local property of the beam, for a certain position along its path, but a property of the beam as a whole. An easy way to change the beam divergence is to change the numerical aperture of the optical system. In practice, this can be achieved by adjusting the diameter of an iris correctly placed in a Fourier plane (accessible at the focal plane of a lens prior to exciting the resin). Closing the iris results in elongating the beam along the optical axis, meaning reducing its divergence but it comes at a price. Cutting off the high spatial frequencies of the beam not only does affect the resolution at the focal plane but also reduces the amount of light available for photopolymerization.

For printing, low beam divergence is essential as shown in refs. [1,2]. This is especially true for printing large objects, but the optical resolution and the overall light budget linked to the printing speed are also very important and a compromise has therefore to be found.

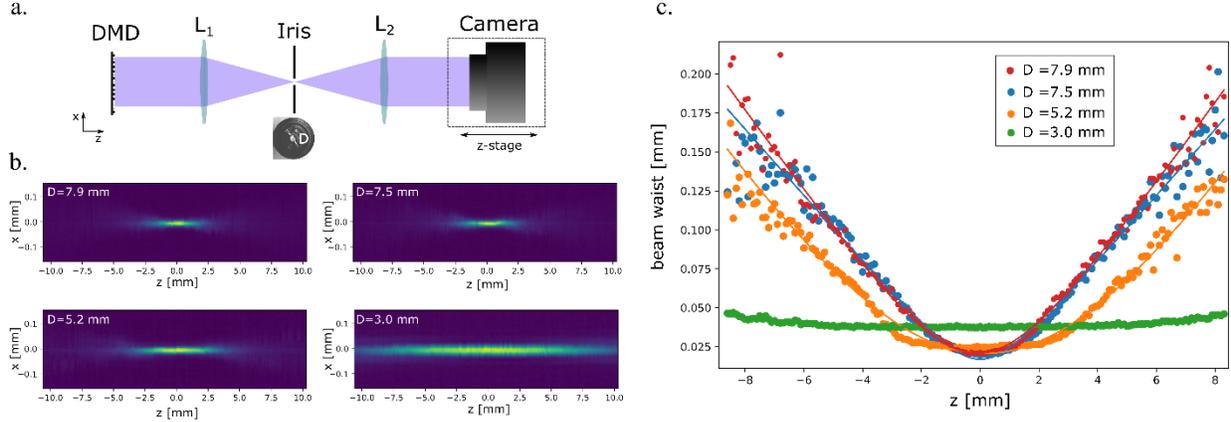

*Figure S3: Effect of four different apertures on the beam divergence. **a.** Schematic of the optical setup used to characterize the beam divergence. **b.** Point-spread-functions (PSF) for four different diameters of the iris D (see inset). **c.** Corresponding beam waist w(z) as a function of z.*

### S4. Beam Divergence simulation with a 3D Gaussian kernel

Most algorithms for tomography assume a perfectly collimated beam of light, but in practice, this is not exactly the case because of light diffraction. Assuming that light propagates in a homogeneous medium, some amount of divergence is unavoidable due to the general nature of waves. That amount is directly related to the numerical aperture of the optical system and is larger for tightly focused beams, corresponding to high NA. Light divergence is one of the factors that limits the resolution with which a three-dimensional dose distribution can be created inside the photosensitive material. To study this effect, we simulate the beam divergence in our Radon transform algorithm. The effect of divergence in tomographic VAM has already been investigated numerically but only in two dimensions, in planes orthogonal to the rotation axis[3]. Here we aim at mimicking the divergence with a finer model by implementing a convolution with a 3D Gaussian Kernel. The parameters of the kernel are adjusted at different depths in order to follow the theoretical trend of the beam waist, $w(z) = w_0 \sqrt{1 + \left(\frac{z}{z_R}\right)^2}$, see Figure S2.b. This results in a global loss of fidelity as can be seen in Figure S2.c. We can also see that it is responsible for a non-homogeneous printing resolution; the voxel size is increasingly larger when moving further away from the imaging plane (middle of the cylindrical vial).

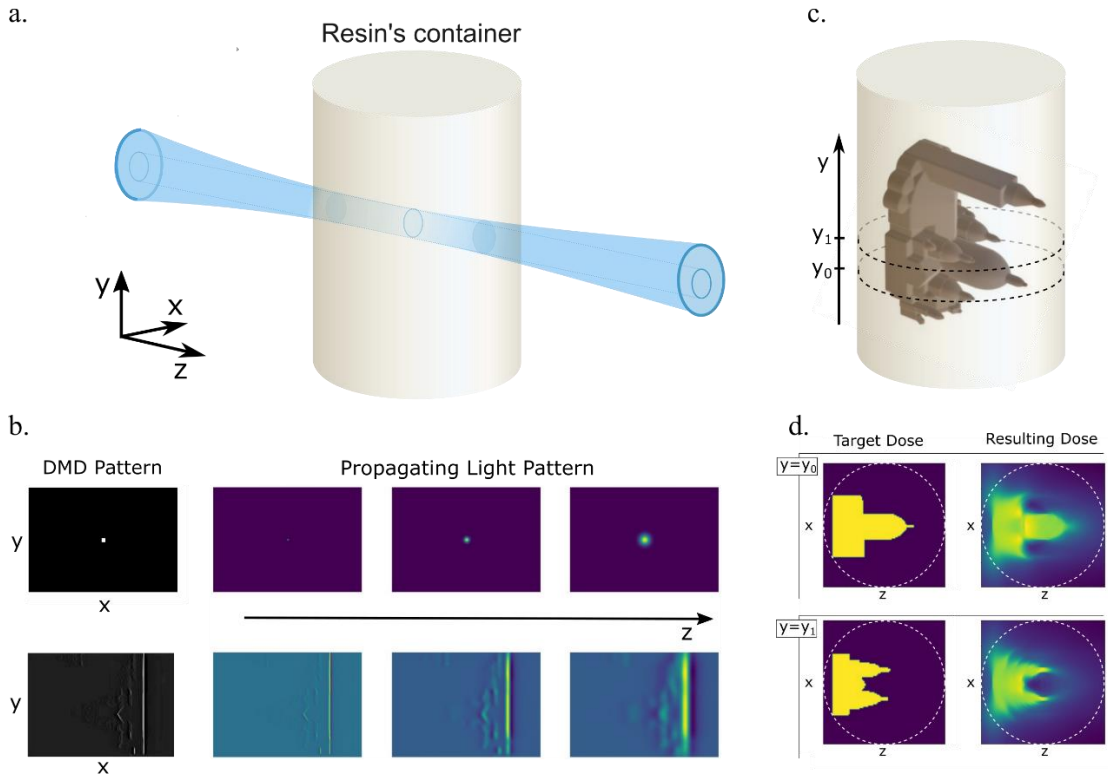

*Figure S4: Simulation of a tomographic printer with a divergent beam. **a.** Schematic illustration of diffracted beam of light across a cylindrical containing the photoresin to be printed. **b.** Simulation of divergence. Top. Only the central pixel of the DMD is in its ON state. Divergence is simulated using a 3D Gaussian kernel whose size is depth dependent (i.e., a function of z). Bottom effect on a 2D DMD pattern projected at a given angle for printing the Sacré Coeur. **c.** Sacré Coeur basilica 3D model inside the cylindrical vial. **d.** Target doses at two different planes and simulating dose obtained for a given divergence. As expected (for a focused beam in the middle of the vial) the effect of divergence is stronger on the vial's edge.*

## S5. Study of Beam divergence on the resulting dose

Using the simulation code described in Section S2, we investigated the effect of the beam divergence on the resulting dose deposited inside the build volume. Two different configurations are considered: on the one hand the DMD is on-axis and illuminates the whole volume of photoresin (as in conventional tomographic VAM) and on the other hand, the DMD is off-centered as in VHAM. As a result, the way the DMD is positioned with respect to the cylindrical vial does not change much the obtained dose. In both cases, the divergence results in a global blur of the light dose. The effect is stronger on the edge as the voxel size due to divergence is larger. The beam divergence is modified by adjusting the Rayleigh range $z_R$ of the optical beam. Short $z_R$ corresponds to high beam divergence.

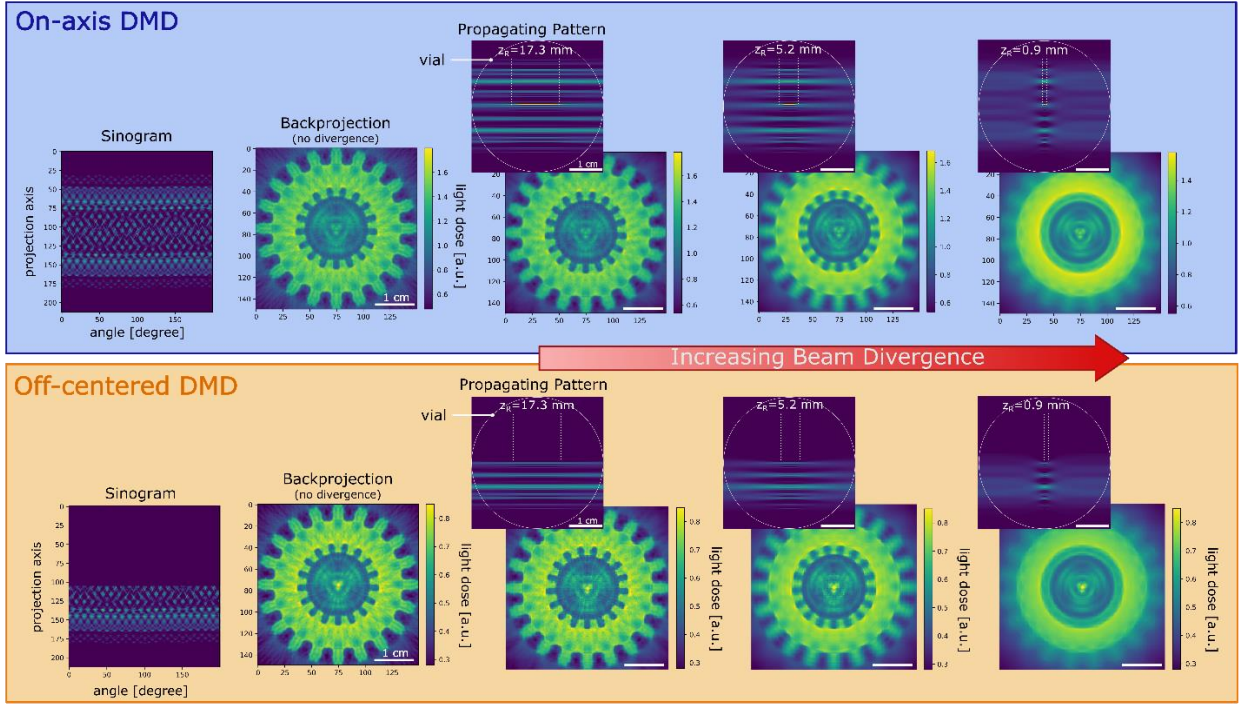

*Figure S5: **Study of Beam divergence on the resulting dose.** From left to right: Sinogram, resulting dose in the case of a straight beam, effect of an increasing beam divergence (i.e., decreasing $z_R$). Inset: Example of a pattern at a given angle. Two different alignments for the DMD are reported, whether on-axis (Top) or off-centered (bottom).*

### S6. Study of defocusing of divergent beams on the resulting dose

Using the simulation code described in Section S2, we investigated the effect of a defocused divergent beam on the resulting dose deposited inside the build volume. As in Section S3 two different configurations are considered: on the one hand the DMD is on-axis and illuminates the whole volume of photoresin (as in conventional tomographic VAM) and on the other hand the DMD is off-centered as in VHAM. This time the way the DMD is positioned with respect to the cylindrical vial radically changes the obtained dose. Whereas the divergence results in a global blur of the light dose, a defocus in the case where only half the vial is illuminated translates into aterfacts/drifts. The effect is stronger on the edge as the voxel size due to divergence is larger, hence the effect of defocusing magnified. The beam defocusing is modified by adjusting position of the focal plane with respect to the rotation axis of the vial.

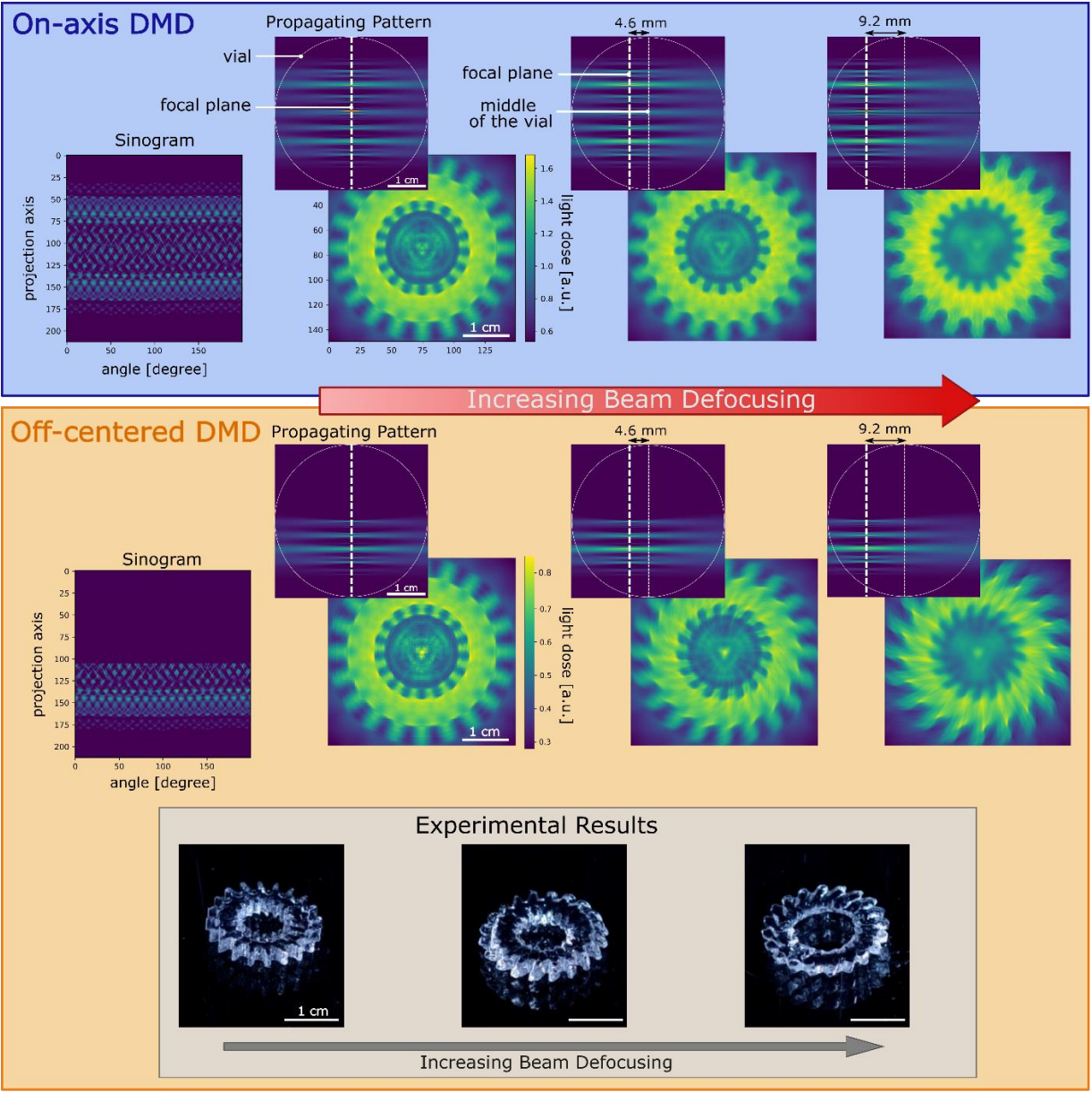

*Figure S6: Study of Beam defocusing on the resulting dose. From left to right: Sinogram, resulting dose in the case of a straight beam, effect of an increasing beam defocusing. Inset: Example of a pattern at a given angle. Two different alignments for the DMD are reported, whether on-axis (Top) or off-centered (bottom). Experimental results obtained with an off-centered DMD confirming the drift for the teeth, observed in the numerical simulations.*

## S.7 Comparison with other volumetric additive manufacturing methods

| Features | Xolography | CAL | | Tomographic VAM | Helical VAM |
|---|---|---|---|---|---|
| DLP chip ($N_x \times N_y$ px) | 3840 × 2160 px | 912 × 1140 px | 2716 × 1528 px | 1024 × 768 px | 1024 × 768 px |
| Image pixel pitch | 21 μm × 21 μm | 38 μm × 38 μm | 31 μm × 31 μm | 23 μm × 23 μm | 23 μm × 23 μm |
| Number of voxels | $N_x \times N_y \times N_z$ | $\frac{\pi}{4}N_x^2 \times N_y$ | $\frac{\pi}{4}N_x^2 \times N_y$ | $\frac{\pi}{4}N_x^2 \times N_y$ | $4\alpha\frac{\pi}{4}N_x^2 \times N_y$ |
| ON pixel intensity | 215 mW.cm$^{-2}$ | 8.7 mW.cm$^{-2}$ | 10.5 mW.cm$^{-2}$ | 400 mW.cm$^{-2}$ | 70 mW.cm$^{-2}$ |
| Print resolution corresp. object's size | ∼ 80 μm 0.9 cm × 0.9 cm × 0.9 cm | ∼ 300 μm 1 cm × 1 cm × 1 cm | - | ∼ 80 μm 1 cm × 1 cm × 1.3 cm | ∼ 200 μm 2.5 cm × 2.5 cm × 3 cm |
| Max. dimensions | 3 cm × 3 cm × 4 cm | 4 cm × 4 cm × 4 cm | 2 cm × 2 cm × 2.7 cm | 1.7 cm × 1.7 cm × 2.3 cm | 3 cm × 3 cm × 6 cm |
| Print time | 1-8 min | 30-120 s | 150 s | 30-120 s | 5-10 min |
| Print speed | 55 mm$^3$.s$^{-1}$ | - | - | 30 mm$^3$.s$^{-1}$ | 75 mm$^3$.s$^{-1}$ |

*Table 1. Comparison of some reported figures of merit of recent light-based volumetric additive manufacturing methods.*